\documentclass[12pt]{wlscirep}
\usepackage[utf8]{inputenc}
\usepackage[T1]{fontenc}
\usepackage{hyperref}
\usepackage{bbold}
\usepackage{subcaption}
\usepackage{setspace}
\usepackage{amsmath}
\usepackage{physics}
\usepackage{mathtools}
\usepackage{nameref}
\usepackage{float}

\title{The Radical Pair Mechanism Cannot Explain  Telecommunication Frequency Effects on Reactive Oxygen Species}

\author[1,2,3,*]{Owaiss Talbi}
\author[1,2,3,*]{Hadi Zadeh-Haghighi}
\author[1,2,3,*]{Christoph Simon}
\affil[1]{Department of Physics and Astronomy, University of Calgary, Calgary, AB T2N 1N4, Canada}
\affil[2]{Institute for Quantum Science and Technology, University of Calgary, Calgary, AB T2N 1N4, Canada}
\affil[3]{Hotchkiss Brain Institute, University of Calgary, Calgary, AB T2N 1N4, Canada}
\affil[*]{Email: \href{mailto:owaiss.talbi@ucalgary.ca}{owaiss.talbi@ucalgary.ca}, \href{mailto:hadi.zadehhaghighi@ucalgary.ca}{hadi.zadehhaghighi@ucalgary.ca}, \href{mailto:csimo@ucalgary.ca}{csimo@ucalgary.ca}}

\begin{abstract}
In order to investigate whether the radical pair mechanism (RPM) can explain the effects of telecommunication frequency radiation on reactive oxygen species production, we modelled the effects of oscillating magnetic fields on radical pair systems. Our analysis indicates that the RPM cannot account for the biological effects observed under exposure to telecommunication frequencies due to negligible effects under low-amplitude conditions used in experimental setups. Observable effects on radical pairs at these frequencies would require hyperfine coupling constants that are precisely fine-tuned to large values that far exceed those naturally occurring within biological systems. We conclude that some other mechanism must be responsible for the effects of telecommunication frequency fields in biological systems.
\end{abstract}
\begin{document}

\flushbottom
\maketitle

\thispagestyle{empty}
\section*{Introduction}

Reactive oxygen species (ROS) are crucial signaling molecules within cells, playing vital roles in various physiological processes \cite{Terzi2020, Sies2020, Sies2017, Gurhan2023}. However, excessive accumulation of ROS can trigger oxidative stress, causing substantial damage to lipids, proteins, and DNA \cite{Imlay2013}. This damage compromises cell function and is associated with the onset of various pathologies, including cancer, cardiovascular diseases, and neurodegenerative conditions \cite{Jackson2009, Handy2017, Incalza2018, Munzel2017, Tarafdar2018, Sbodio2019}. Thus, the regulation of ROS levels is essential for maintaining cellular integrity and preventing the progression of these pathologies \cite{Sies2020}.

Experimental evidence, detailed in Table \ref{tab:emr_bioeffects}, shows that electromagnetic radiation at telecommunication frequencies, particularly within the ultra-high frequency (UHF) range, can influence ROS levels even at low amplitudes. Studies utilizing exposure to UHF electromagnetic fields have observed changes in ROS production within various cell types.
For instance, Luukkonen et al. reported that exposure to 872 MHz radiation at a specific absorption rate (SAR) of 5 W/kg induced ROS production and DNA damage in human SH-SY5Y neuroblastoma cells \cite{Luukkonen2009}. Similarly, other investigations have shown increased ROS levels in human peripheral blood mononuclear cells \cite{Kazemi2015}, human HEK293 cells \cite{Pooam2022}, and astrocytes \cite{Campisi2010} following exposure at frequencies relevant to mobile communication technologies. Moreover, the effects extend beyond cellular responses to affect whole organisms. Research has shown that Drosophila exposed to DECT (Digital Enhanced Cordless Telephone) base radiation 1.88-1.90 GHz exhibited elevated ROS levels in both bodies and ovaries \cite{Manta2014}. In addition, studies on human spermatozoa exposed to 2.45 GHz Wi-Fi radiation revealed significant oxidative stress damage, including increased ROS levels, DNA fragmentation, and decreased sperm motility and vitality \cite{Ding2018}. 

\begin{table}[htbp]
    \centering
    \caption{Summary of ultra-high frequency effects on ROS.}
    \label{tab:emr_bioeffects}
    \begin{tabular}{p{2.5cm}p{11cm}l}  
    \hline
    \centering
     \textbf{frequency} &  \textbf{Bioeffect} &  \textbf{Ref} \\ \hline
    \centering
     900 MHz &  Increased ROS level in human SH-SY5Y cells &  \cite{Luukkonen2009} \\ \hline 
    \centering  872 MHz &  Increased ROS level in human peripheral blood mononuclear cells &  \cite{Kazemi2015} \\ 
    \hline 
    \centering  1.8 GHz &  Increased ROS level in human HEK293 cells &  \cite{Pooam2022} \\ \hline
    \centering  900 MHz &  Increased ROS level and DNA fragmentation in astrocytes &  \cite{Campisi2010} \\ \hline  
    \centering  1.88-1.90 GHz &  Increased ROS level in Drosophila &  \cite{Manta2014}\\
    \hline 
    \centering  2.45 GHz &  Increased ROS level in human semen &  \cite{Ding2018}\\
    \hline
    \centering  870 MHz &  Increased ROS level in human ejaculated semen &  \cite{Agarwal2009}\\
    \hline 
    \centering  1.8 GHz &  Increased ROS level and DNA damage in human spermatzoa &  \cite{DeIuliis2009}\\
    \hline 
    \centering  940 MHz &  Increased ROS level in HEK cells &  \cite{Sefidbakht2014}\\
    \hline 
    \end{tabular}
\end{table}

Weak magnetic fields (MFs) have been known to influence chemical reactions. Numerous studies have documented the effects of static magnetic fields (SMFs) on ROS production in various systems, highlighting their biological significance. For instance, Calabrò et al. reported that exposure to a static magnetic field of 2.2 mT significantly increased ROS production in human SH-SY5Y neuroblastoma cells \cite{Calabro2013}. Bekhite et al. found that static magnetic fields ranging from 0.2 mT to 5 mT elevated ROS levels in mouse embryoid bodies, indicating the role of ROS in SMF-induced differentiation processes \cite{Bekhite2013}. Additionally, Martino et al. demonstrated that low-level magnetic fields (45 $\mu$T to 60 $\mu$T) modulated hydrogen peroxide (\( \text{H}_2 \text{O}_2 \)) production in various cell types, including cancer cells and endothelial cells, suggesting a broader impact of magnetic fields on cellular redox states \cite{Martino2011}.

The radical pair mechanism (RPM) can explain static magnetic field effects on chemical reactions, originating from spin chemistry, with magnetoreception in avian species being a notable example \cite{hayashi2004introduction, Hochstoeger2020, Huelga2013, Johnsen2005, Wiltschko2023, Ritz2000, Ritz2004, Zadeh-Haghighi2022, Maeda2008, Kerpal2019, Mouritsen2022, Xu2021}. This model is based on the creation of radical pairs—molecules with unpaired electron spins in the presence of nearby nuclear spins—that are exquisitely sensitive to weak MFs \cite{Timmel1996, Hore2012, Hore2020}. Recent studies have illustrated how RPM can influence ROS production in live cells. Usselman et al. demonstrated that coherent electron spin dynamics in radical pairs formed at flavoenzyme centers can modulate ROS levels through singlet-triplet interconversion, impacted by both static and oscillating magnetic fields \cite{Usselman2016}. Superoxide radicals (\( \text{O}_2^{\boldsymbol{\cdot} -} \)), as primary forms of ROS, are generated through two main pathways: mitochondrial electron transport and the action of NADPH oxidase enzymes \cite{Bedard2007, Liu2002, Terzi2020, Moghadam2021, Murphy2009, HernansanzAgustin2021}. This electron transfer process—and thus ROS generation—can be modulated by magnetic fields through RPM \cite{Usselman2014}. Usselman et al. further showed that these fields can alter the cellular balance of \( \text{O}_2^{\boldsymbol{\cdot} -} \) and \( \text{H}_2 \text{O}_2 \), indicating a broader spectrum of mechanisms at play \cite{Usselman2016}. This is further supported by the observation that both the mitochondrial electron transport chain and Nox enzymes can generate magnetically sensitive flavin and superoxide-based radical pairs, implicating them as potential contributors to the magnetic field effects on ROS levels \cite{Rishabh2022, Usselman2016}.

Given the established effects of static magnetic fields on ROS through RPM, it is pertinent to investigate whether the observed effects of telecommunication frequency radiation on ROS could be due to oscillating magnetic fields (OMFs) in the context of the RPM. Our investigation shows that the RPM cannot account for the observed effects of radio frequencies typically found within telecommunication devices on ROS levels. This suggests the presence of another mechanism, possibly involving the electric component of the UHF fields.

\section*{Results}
\subsection*{Radical Pair Mechanism}
Radical pairs, formed during processes involving molecular bond breakage or electron transfer events, exhibit behavior dictated by their intrinsic angular momentum, represented by the spin quantum number (\(\mathrm{S}\)) and the spin projection quantum number (\(m_s\)) \cite{Zadeh-Haghighi2022}. This intrinsic angular momentum facilitates interactions with external magnetic fields and adjacent spins. Importantly, the conservation of angular momentum during these interactions means that radical pairs tend to align according to the spin state of their precursor molecules \cite{Zadeh-Haghighi2022, Gerson2003, efimova2008role}. Consequently, the interaction of two unpaired electron spins can yield either a singlet state \(|\mathrm{S}\rangle\) (\(\mathrm{S} = 0\), \(m_s = 0\)) or triplet states \(|\mathrm{T}\rangle\) (\(\mathrm{S} = 1\), \(m_s = 0, \pm1\)). The initial state of a radical pair, whether in a singlet or triplet configuration, is influenced by the surrounding nuclear spins, which are in a maximally mixed state due to thermalization. This state reflects an equilibrium distribution of spin states at radical pair formation, described by a density matrix proportional to the identity matrix, signifying an ensemble of spin states without coherence \cite{Luo2022}.

\vspace*{12pt}
\noindent \textit{Hyperfine and Zeeman interactions.} Radical pair dynamics are greatly influenced by Zeeman and hyperfine interactions. The Zeeman interaction aligns electron spins with an external magnetic field, altering energy states and influencing reaction pathways \cite{improta2004interplay}. This interaction is described by the following:
\begin{equation}
\hat{H}_{\text{Zeeman}} = -\gamma_e \hat{\mathbf{S}} \cdot \mathbf{B},
\end{equation}
where $\gamma_e$ represents the electron gyromagnetic ratio, $\hat{\mathbf{S}}$ the electron spin operator, and $\mathbf{B}$ is the external magnetic field \cite{Zadeh-Haghighi2022, improta2004interplay}. While electron spins are significantly affected by these fields, the much smaller nuclear gyromagnetic ratios mean that the direct influence of magnetic fields on nuclear spins within radical molecules is negligible for radical pair dynamics \cite{Zadeh-Haghighi2022}.

The hyperfine interaction, which encompasses both isotropic and anisotropic contributions, couples electron spins with nuclear spins. The anisotropic part, akin to a dipole-dipole interaction, depends on the spatial distribution of the electron. It is often assumed to be averaged out to zero due to molecular motion. This assumption is based on the premise of rapid molecular tumbling in solution \cite{atkins2005molecular, Player, Hogben}. Conversely, the isotropic component, known as the Fermi contact interaction, emerges from the direct interaction within the nucleus itself, playing a significant role in singlet-triplet transitions of radical pairs \cite{atkins2005molecular, Player, Hogben}. This interaction is described as the following:
\begin{equation}
\hat{H}_{\text{HFI}} = a_i \hat{\mathbf{S}} \cdot \hat{\mathbf{I}}_i,
\end{equation}
where $\hat{\mathbf{I}}_i$ represents the nuclear spin of the $i$-th nucleus and $a_i$ the isotropic hyperfine coupling constant (HFCC). Additionally, the exchange and dipolar interactions also contribute to the dynamics of radical pairs. The exchange interaction, which accounts for the indistinguishability of electrons, is assumed to weaken exponentially as the radical pair separates \cite{Zadeh-Haghighi2022}. The dipolar interaction arises from the magnetic moments in radical pairs and along with exchange interaction, they can inhibit singlet-triplet interconversion \cite{efimova2008role, kattnig2017sensitivity, nohr2017determination, babcock2021radical}. However, in biological contexts where radical pairs are well-separated, the effects of both exchange and dipolar interactions are usually minimal \cite{kattnig2017sensitivity, nohr2017determination}. 

\vspace*{12pt}
\noindent \textit{Spin dynamics and radical pair interconversion.}
The oscillation between the singlet and triplet states of radical pairs, known as quantum beats, plays a crucial role in this dynamic \cite{Anisimov1983}. These oscillations become pronounced when radicals are spatially separated enough to allow for coherent interconversion \cite{Zadeh-Haghighi2022, Hore2021, Hore2016}. 
Magnetic fields, through hyperfine and Zeeman interactions, finely tune these conversions. At lower field strengths, hyperfine interactions predominantly drive the singlet-triplet interconversion, dictating the chemical pathways the radicals may follow based on the coherence of their spin dynamics\cite{Zadeh-Haghighi2022, Hore2021}. The coherence in spin dynamics and the interconversion between singlet and triplet states can be mathematically described by the Liouville-von Neumann equation, which governs the time evolution of the spin density matrix, $\hat{\rho}$, incorporating coherent superpositions and external perturbations like magnetic field effects  \cite{Zadeh-Haghighi2022, Timmel1998}. It is given by:
\begin{equation}
\frac{d\hat{\rho}(t)}{dt} = -\frac{i}{\hbar}[\hat{H}, \hat{\rho}(t)],
\end{equation}
where $\hat{H}$ represents the Hamiltonian of the system including all interactions such as the Zeeman effect and hyperfine coupling \cite{Player,Hogben,Zadeh-Haghighi2022,Zadeh-Haghighi2023}.

The initial state of a radical pair, considering nuclear spins in thermal equilibrium, is described by a spin density matrix proportional to the identity matrix. For radical pairs initially in singlet state, the spin density matrix is defined as:
\begin{equation}
    \hat{\rho}(0) = \frac{1}{M} \hat{P}_{\mathrm{S}} = \frac{1}{M} |\mathrm{S}\rangle \langle \mathrm{S}| \otimes \mathbb{\hat{1}}_{M},
\end{equation}
where $\hat{P}_S$ is the projection operator for the singlet state, $\mathbb{\hat{1}}_M$ represents the identity matrix, and \(M\) is the nuclear spin multiplicity, defined as \(M = \prod_{i}(2I_i + 1)\), with \(I_i\) being the spin angular momentum of the \(i\)-th nucleus \cite{Luo2022, Zadeh-Haghighi2022}.

The probability of finding a radical pair in a singlet state at any given time is fundamental in the study of the spin dynamics. These probabilities are determined by the Hamiltonian of the system through the solution of the Liouville-von Neumann equation. To quantify these, we consider the trace over the product of the singlet projection operator $\hat{P}_S$ with the spin density matrix $\hat{\rho}(t)$:
\begin{equation}
\langle \hat{P}_S(t) \rangle = \text{Tr}[\hat{P}_S \hat{\rho}(t)],
\end{equation}
where $\langle \hat{P}_S(t) \rangle$ represents the probability of the radical pair being in the singlet state at time $t$ \cite{Zadeh-Haghighi2022, Player, Hogben, Timmel1998}.

To incorporate real-world perturbations, spin relaxation can be added phenomenologically into the spin dynamics equation. This adjustment simulates the exponential decay towards equilibrium—characterized by a 25\% probability for the singlet state and a 75\% probability distributed among the triplet states, guided by the spin relaxation rate \( r \). The adjusted singlet probability is given by:
\begin{equation}
    \langle \hat{P}_S(t) \rangle \rightarrow \frac{1}{4} + \left( \langle \hat{P}_S(t) \rangle - \frac{1}{4} \right) e^{-rt},
\end{equation}
this modification captures how random molecular motions modulate electron spins, causing a shift towards the equilibrium distribution of 25\% singlet and 75\% triplet states as time progresses \cite{Zadeh-Haghighi2022, Hore2019}.

Following the insights from Timmel et al., the chemical fate of the radical pair is elucidated by considering separate spin-selective reactions for singlet and triplet pairs. These reactions are modelled with identical first-order rate constants, $k$ \cite{Timmel1998}. The yield of the singlet state, denoted as $\Phi_S$ is characterized by the following expression:
\begin{equation}
\Phi_S = k \int_{0}^{\infty} \langle \hat{P}_S(t) \rangle e^{-kt} dt,
\end{equation}
integrating over the modified probabilities that account for spin relaxation. Since $\Phi_S + \Phi_T = 1$, knowledge of one yield allows for the calculation of the other, encapsulating the complete dynamics of the radical pair's chemical fate \cite{Zadeh-Haghighi2022}.

\subsection*{Specific Absorption Rate}

The specific absorption rate is a key metric in assessing how electromagnetic fields affect biological tissues, particularly under radio frequency (RF) exposures. SAR measures the energy absorbed per unit mass of tissue, given in watts per kilogram (W/kg), and is vital for understanding the potential health impacts of RF radiation\cite{Poljak2018}. SAR provides a standardized way to compare exposure levels across different RF sources, making it important in regulatory settings. Experiments that investigate radio frequency effects on biological systems often report SAR values to quantify the intensity of exposure. For example, SAR data helps correlate RF exposure levels with biological outcomes like increased ROS production or DNA damage. To estimate the magnetic fields involved in these experiments, we use a conversion formula that relates SAR to magnetic field strengths. This estimation is crucial for aligning theoretical models with experimental conditions. The formula is as follows:
 \begin{equation}
 \text{SAR} = \frac{\sigma |E|^2}{2\rho},
 \end{equation}
where \(\sigma\) is the electrical conductivity of the tissue in S/m, \(E\) is the electric field in medium in V/m, and \(\rho\) is the mass density of the tissue in kg/m\(^3\) \cite{Poljak2018}.

Given the experiment involving SH-SY5Y cells, where the SAR, dielectric constant ($\epsilon_r$), effective electric conductivity ($\sigma$), and sample density ($\rho$) are specified, we seek to estimate the magnetic field strength ($B$) in Tesla within the biological medium \cite{Buttiglione2007}. These values have been applied as representative approximations for other studies lacking such detailed data, allowing for a consistent approach to estimating magnetic field strengths across various experimental setups. Assuming that the magnetic permeability of biological tissues approximates that of free space ($\mu_0$), and considering the relative permittivity ($\epsilon_r = 75$), conductivity ($\sigma$ = 1.9 \,\text{S/m}), and density ($\rho = 10^3 \, \text{kg/m}^3$), we can calculate the magnetic field from the electric field in the medium by recognizing that the magnetic field ($B$) and magnetic field intensity ($H$) are related by:
\begin{equation}
 B = \mu_0 H,
\end{equation}
where $\mu_0$ is the magnetic permeability of free space ($4\pi \times 10^{-7} \, \text{H/m}$). Next, considering the impedance ($Z$) of the medium, the magnetic field intensity is related to the electric field by:
\begin{equation}
 H = \frac{E}{Z},
\end{equation}
where $Z$, the impedance of the medium, incorporates the medium's conductivity and is given by:
\begin{equation}
 Z = \sqrt{\frac{j\omega\mu_0}{\sigma + j\omega\epsilon_0\epsilon_r}},
\end{equation}
in which $\epsilon_0$ is the permittivity of free space ($8.85 \times 10^{-12} \, \text{F/m}$), $j$ is the imaginary unit, and $\omega$ is the angular frequency of the RF field \cite{Wolski}. This method enables the estimation of the magnetic from known values of $E$, $\epsilon_r$, $\sigma$, and $\rho$, factoring in the medium's electrical properties \cite{Poljak2018,Buttiglione2007, Wolski}. Using this method, we estimated range of magnetic field amplitudes used in experimental conditions extends from approximately 95 nT to 5 $\mu$T. For instance, in the study by Manta et al. \cite{Manta2014}, the calculated SAR value of 0.009 W/kg corresponds to an estimated magnetic field amplitude of approximately 95 nT. Conversely, in the study by De Iuliis et al. \cite{DeIuliis2009}, the SAR values ranging from 0.4 W/kg to 27.5 W/kg correspond to an estimated upper magnetic field amplitude of approximately 5 $\mu$T. To align our analysis more closely with practical experimental scenarios, we selected a conservative OMF amplitude of $B_1 = 5$ $\mu$T for comparative analysis. This chosen amplitude represents the higher end of magnetic fields estimated from SAR values in experimental settings, ensuring that our conclusions are applicable to all considered exposure scenarios.

\subsection*{Magnetic Field Effects on Radical Pair Dynamics}

With a foundational understanding of the RPM in place, we now turn our attention to the effects of magnetic fields on the dynamics of radical pairs. By applying the principles discussed, we investigate how these fields influence the quantum states and reaction yields of radical pairs.
The Hamiltonian for a typical radical pair system in the presence of a static magnetic field, which encompasses both Zeeman interaction and hyperfine coupling effects, is expressed as:
\begin{equation}
\hat{H} = \omega(\hat{S}_{\mathit{A}_z} + \hat{S}_{B_z}) + \sum_i a_i \hat{S}_A \cdot \hat{I}_i + \sum_j a_j \hat{S}_B \cdot \hat{I}_j,
\end{equation}
where \( \omega \) is the Larmor precession frequency for the electrons, proportional to the strength of the magnetic field and the electron magnetogyric ratio (\( \omega  = -\gamma_e B\)), \( \hat{S}_{A_z} \) and \( \hat{S}_{B_z} \) are the spin operators for electrons A and B, $\hat{I}_i$ and $\hat{I}_j$ represent the nuclear spin operators for nuclei interacting with electrons A and B respectively, and the coefficients $a_i$ and $a_j$ are the isotropic HFCCs \cite{improta2004interplay}. This Hamiltonian allows us to explore the effects of static magnetic fields on the singlet and triplet evolution of radical pairs.

With a focus on ROS, we adopt a radical pair system comprising FADH\(^{\boldsymbol{\cdot}}\) and O\(_2^{\boldsymbol{\cdot -}}\) to explore the magnetic field effects on radical pair dynamics. The correlated spins within the radical pair are modelled as [FADH\(^{\cdot}\) ...O\(_2^{\cdot -}\)], where the reaction produces either H\(_2\)O\(_2\) from the singlet state or maintains O\(_2^{\cdot -}\) from the triplet state, without considering any conversion from O\(_2^{\cdot -}\) to H\(_2\)O\(_2\) \cite{Rishabh2022,Usselman2016}. The unpaired electron on FADH\(^{\cdot}\) is assumed to interact only with its H5 nucleus \cite{Rishabh2022, Lee2014}. In contrast, the unpaired electron on O\(_2^{\cdot -}\) is not engaged in hyperfine interactions due to zero nuclear spin of the superoxide. This choice to consider only the H5 nucleus is justified by the work of Hiscock et al., which highlights the significant impact of hyperfine interactions on radical pair dynamics and resonance effects. Hiscock’s studies show that the presence of multiple nuclei can lead to a variety of energy-level spacings, affecting the resonance conditions and the magnetic sensitivity of the radical pair system \cite{hiscock_disruption_2017}. In our system, the H5 nucleus in FADH\(^{\boldsymbol{\cdot}}\) has a significantly higher hyperfine coupling constant (\(a = 802.9\) $\mu$T) compared to other nuclei, making it the most relevant for our analysis. Including additional nuclei with lower hyperfine coupling constants would complicate the model without substantially altering the primary effects we are investigating. Therefore, our choice to focus on the single H5 nucleus is both a simplification and a reflection of its dominant role in the hyperfine interactions relevant to our study. Note that this choice would likely overestimate the magnetic field effect \cite{Rishabh2022, Lee2014, hiscock_disruption_2017}.

\vspace*{12pt}
\noindent \textit{Oscillating Magnetic Field.} 
Oscillating magnetic fields introduce time-dependent dynamics that can influence radical pair processes. These fields, characterized by their frequency and amplitude, induce changes in the electron spin states over time, which can potentially alter the reactivity and yields of singlet and triplet products. Extending our exploration of radical pair dynamics, we now focus on the effects of OMFs, particularly at radio frequencies, to assess whether the RPM provides a satisfactory explanation for the experimental observations regarding ROS production under such fields. To incorporate the effect of an OMF on the system, the Hamiltonian is represented as:

\begin{equation}
\hat{H} = \sum_j^2 \sum_i^N a_{ji} \hat{S}_j \cdot \hat{I}_i + \omega_0(\hat{S}_{j_x}  \cos{\theta} + \hat{S}_{j_z} \sin{\theta}) - \gamma_e B_1\hat{S}_{j_x}\sin({\omega_{RF} t}),
\end{equation}
where $a_{ji}$ represents the isotropic HFCC between the $j$-th electron spin operator $\hat{S}_j$ and the $i$-th nuclear spin operator $\hat{I}_i$. The term $\omega_0$ accounts for the static component of the magnetic field with $\theta$ indicating the angle between the two magnetic fields, while $\gamma_e B_1\sin({\omega_{RF} t})$ introduces the oscillating magnetic field where $B_1$ is the amplitude of the oscillating field, and $\omega_{RF}$ is its angular frequency \cite{Rodgers2005}.
\begin{figure}[H]
    \centering
    \includegraphics[width=100mm]{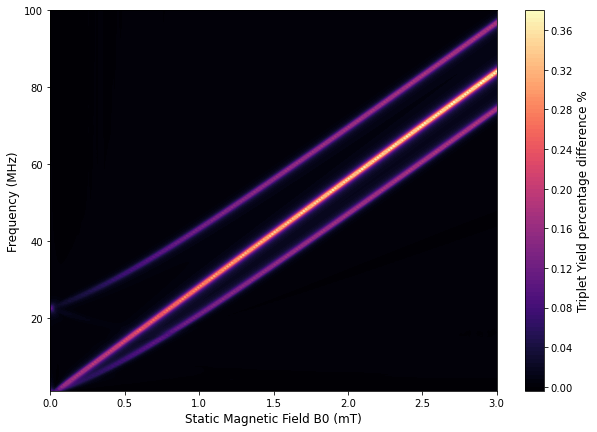}
    \caption{Variation in triplet yield with static magnetic field strength and radio frequency for a constant isotropic HFCC ($a = 802.9 \, \mu\text{T}$), OMF amplitude $B_1 = 5$ $\mu$T, reaction rate \(k = 3 \times 10^6 \, \text{s}^{-1}\), relaxation rate \(r = 1 \times 10^6 \, \text{s}^{-1}\), and spin multiplicity \(M = 2\). Resonant lines indicate enhanced triplet yields at specific field-frequency combinations.}
    \label{Figure1.png}
\end{figure}

For our calculations, we set the reaction rate \(k = 3 \times 10^6 \, \text{s}^{-1}\) and the relaxation rate \(r = 1 \times 10^6 \, \text{s}^{-1}\). Assuming the radical pair initially in the singlet state, Figure \ref{Figure1.png} illustrates how the triplet yield of our radical pair system responds to varying static magnetic field strengths and osciallting magnetic field frequencies, with an isotropic HFCC of $a = 802.9$ $\mu$T. The amplitude of the radio frequency field applied here is $B_1 = 5$ $\mu$T, a value chosen to explore the influence of radio frequencies at a relatively low amplitude on radical pair dynamics compared to the range of varying static magnetic field. 

We observe resonant peaks where the combination of static field strength and radio frequency increases the triplet yield. These conditions, where specific frequencies of the oscillating magnetic field combined with particular static magnetic field strengths, induce noticeable enhancements in triplet yields, demonstrating the potential of an oscillating field to modify the standard dynamics of the radical pair system. At these resonant points, the applied radio frequency enhances the conversion of radical pairs to the triplet state, which diverges from the yields observed in the absence of an OMF, leading to a shift in reaction products \cite{Rishabh2022, Rodgers2005}.

\begin{figure}[htbp]
    \centering
    \begin{subfigure}[b]{0.49\textwidth}
        \centering
        \includegraphics[width=\textwidth]{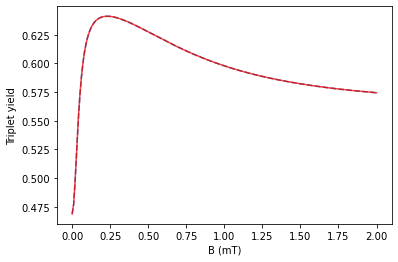}
        \caption{ \centering}
        \label{Figure2_a.png}
    \end{subfigure}
    \hfill
    \begin{subfigure}[b]{0.49\textwidth}
        \centering
        \includegraphics[width=\textwidth]{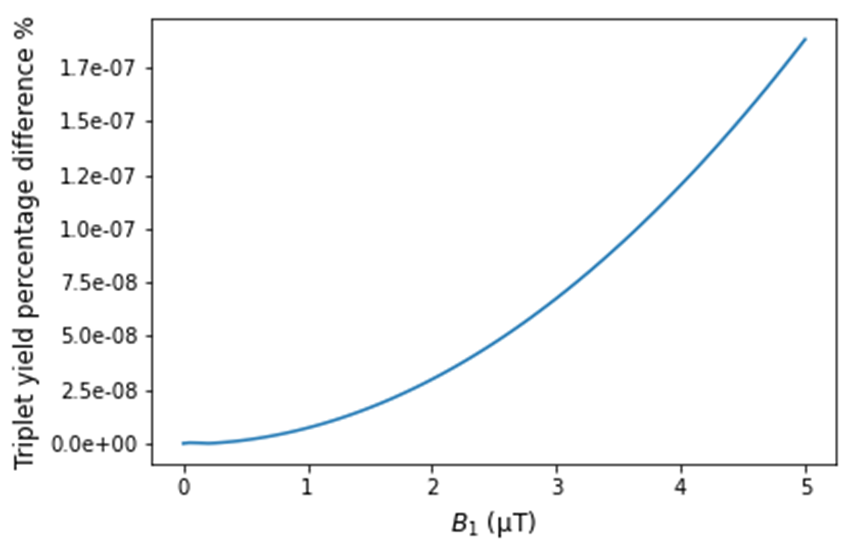}
        \caption{ \centering}
        \label{Figure2_b.png}
    \end{subfigure}
    \caption{ Analysis of radical pair triplet yield at a frequency of 872 MHz. The analysis was conducted with hyperfine coupling constant \(a = 802.9\) $\mu$T, reaction rate \(k = 3 \times 10^6 \, \text{s}^{-1}\), relaxation rate \(r = 1 \times 10^6 \, \text{s}^{-1}\), and spin multiplicity \(M = 2\). \(\mathbf{(a)}\) Comparative analysis of [FADH\(^{\cdot}\) ...O\(_2^{\cdot -}\)] radical pair triplet yield with 872 MHz and an amplitude $B_1 = 5$ $\mu$T (solid red line) and without a radio frequency field (dashed blue line). The overlapping lines show a negligible influence of high-frequency fields at this frequency. \(\mathbf{(b)}\) Effect of varying OMF amplitudes from 0 to 5 $\mu$T on the triplet yield for a system at a static magnetic field of 50 $\mu$T. The results indicate a minimal influence, with approximately 1.88 $\times 10^{-7}$\% difference at 5 $\mu$T OMF amplitude.}
    \label{fig:high_freq_overlap}
\end{figure}

To investigate effects in higher-frequency fields, particularly those above 800 MHz, we studied the influence of the initial hyperfine coupling constant ($a = 802.9$ $\mu$T) and amplitude $B_1 = 5$ $\mu$T. Our observations at 872 MHz reveal that the effects observed when an oscillating magnetic field is applied alongside a static magnetic field are very small compared to when only the static magnetic field is present. As shown in Figure \ref{fig:high_freq_overlap}, the presence of an OMF at 872 MHz produces effects similar to those observed without it. When comparing these scenarios in the presence of a static magnetic field at geomagnetic field strength, the effect size is extremely small, with only an approximate (1.88 $\times$ 10$^{-7}$)\% difference between applying and not applying an OMF at 872 MHz. This observation suggests that, for the given initial HFCC, the impact of UHF fields on the dynamics of the radical pair may not be significant. Such findings prompt further exploration into additional factors that might contribute to the interactions observed at such frequencies.

\begin{figure}[H]
    \centering
    \includegraphics[width=100mm]{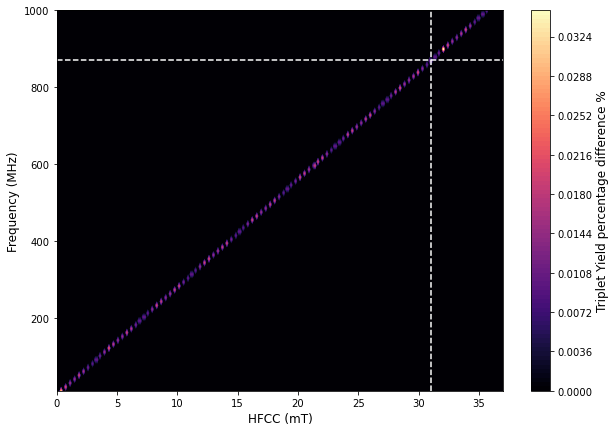}
    \caption{ Variation in triplet yield difference with and without OMF ($B_1 = 5$ $\mu$T) as a function of hyperfine coupling constant (HFCC) and frequency at geomagnetic static field strength ($B_0 = 50$ $\mu$T). The dashed white line at 872 MHz indicates the HFCC (31.14 mT) needed for a measurable OMF effect. The analysis was conducted with reaction rate \(k = 3 \times 10^6 \, \text{s}^{-1}\), relaxation rate \(r = 1 \times 10^6 \, \text{s}^{-1}\), and spin multiplicity \(M = 2\).}
    \label{Figure3.png}
\end{figure}

The hyperfine coupling constant is a critical factor that modulates the response of radical pair systems to magnetic fields. It indicates a possible mechanism through which higher frequency fields can exert biological effects even when the static magnetic field aligns with Earth's geomagnetic field strengths. Figure \ref{Figure3.png} addresses how varying the HFCC impact the triplet yield of a radical pair under a static magnetic field typical of the geomagnetic range.

\begin{figure}[H]
\centering
\includegraphics[width=100mm]{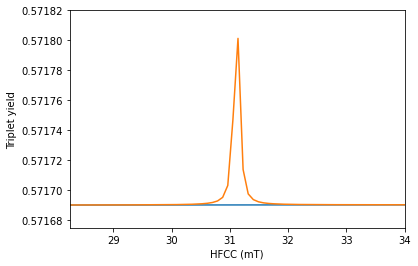}
\caption{ Triplet yield as a function of the hyperfine coupling constant (HFCC) at a geomagnetic-strength static field ($B_0 = 50$ $\mu$T) and a radio frequency of 872 MHz. The blue line indicates the triplet yield in the absence of an OMF, whereas the orange line depicts the yield when subjected to an OMF with strength $B_1 = 5$ $\mu$T. The graph highlights a narrow HFCC range where the OMF effect is maximized, around an HFCC value of 31.14 mT. The analysis was conducted with reaction rate \(k = 3 \times 10^6 \, \text{s}^{-1}\), relaxation rate \(r = 1 \times 10^6 \, \text{s}^{-1}\), and spin multiplicity \(M = 2\).}
\label{Figure4.png}
\end{figure}

Our analysis reveals a direct correlation between the magnitude of the HFCC and the effect of an oscillating magnetic field on the triplet yield at higher frequencies. Notably, an HFCC of approximately 31.14 mT is required to observe an effect at a radio frequency of 872 MHz, shown as dashed white line on Figure \ref{Figure3.png}. Figure \ref{Figure4.png} further illustrates that the triplet yield varies significantly with HFCC, showing a narrow range around 31.14 mT where the effect of the oscillating magnetic field is maximized. This suggests that not only would the HFCC need to be exceptionally large but also finely tuned to account for effects at such high frequencies. This seems highly improbable, casting doubt on the likelihood that large HFCC values are the correct explanation for the observed phenomena, as typical HFCC values in biological systems are much lower, generally in the range of microtesla to a few millitesla. For instance, the HFCC for hydrogen nuclei in biological radicals can be around 7.45 mT for serine oxyradicals and 1.86 mT for tyrosine oxyradicals \cite{Nair2024}. Other studies on radicals derived from amino acids and other biological molecules show that the HFCCs typically range up to 5 mT \cite{Ban2013}.

\begin{figure}[htbp]
    \centering
    \includegraphics[width=100mm]{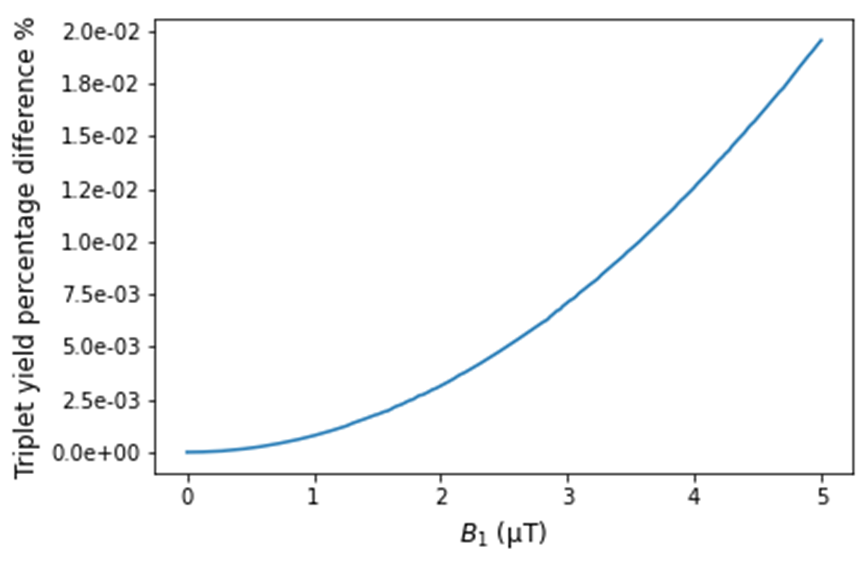}
    \caption{ Effect of varying OMF amplitudes from 0 to 5 $\mu$T on the triplet yield for a system at a static magnetic field of 50 $\mu$T for HFCC $a = 31.14$ mT. The results indicate a minimal influence of a field at a frequency of 872 MHz for the adjusted HFCC, with approximately 0.02\% difference at 5 $\mu$T OMF amplitude. The analysis was conducted with a reaction rate \(k = 3 \times 10^6 \, \text{s}^{-1}\), relaxation rate \(r = 1 \times 10^6 \, \text{s}^{-1}\), and spin multiplicity \(M = 2\).}
    \label{Figure5.png}
\end{figure}

Nevertheless, given the minimal impact observed with the initial HFCC, we adjusted the HFCC to 31.14 mT to explore the influence of radio frequency on the triplet yield within a geomagnetic field range. We kept the static magnetic field constant at 50 $\mu$T to simulate the geomagnetic field and varied the OMF amplitude from 0 to 5 $\mu$T. With this adjustment, we observed the effects of the UHF fields more clearly. Specifically, for this large and fine-tuned HFCC, the percentage difference in triplet yield between applying and not applying 5 $\mu$T OMF to the system is approximately 0.02\%, as illustrated in Figure \ref{Figure5.png}.

\begin{figure}[H]
\centering
\includegraphics[width=100mm]{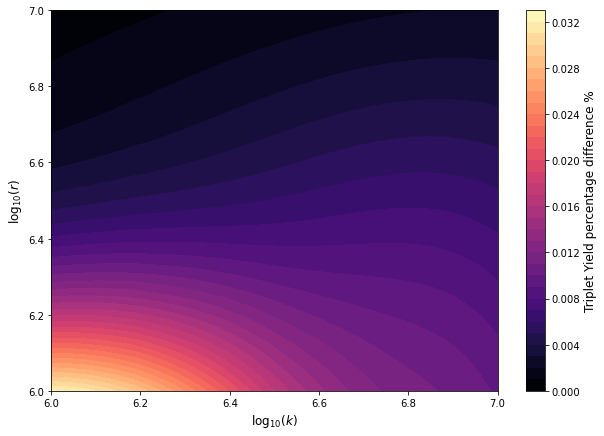}
\caption{Analysis showing the percentage difference in triplet yield with and without an OMF at an amplitude of $B_1 = 5$ $\mu$T, frequency of 872 MHz, HFCC $a_1 = 31.14$ mT, and spin multiplicity \(M = 2\). across various relaxation ($r$) and reaction ($k$) rates at a presence of geomagnetic field ($B_0 = 50$ $\mu$T).}
\label{Figure6.png}
\end{figure}

Furthermore, we investigated the effects of varying reaction and relaxation rates while keeping the OMF amplitude fixed at 5 $\mu$T and the static magnetic field constant at 50 $\mu$T. Figure \ref{Figure6.png} shows that, even for this very large and finely tuned value of the HFCC, the influence of the magnetic component of the electromagnetic radiation at telecommunication frequencies remains less than 0.033\%, underscoring the limited impact of low amplitude OMF on the radical pair system.

\section*{Discussion}
Our exploration of the radical pair mechanism sought to understand the effects of telecommunication frequency radiation on the production of reactive oxygen species  within biological systems. Our study uncovers limitations of the radical pair mechanism in explaining the observed effects of radio frequency fields, especially those in the telecommunications frequency range, such as 872 MHz. A critical finding of our analysis is the requirement for hyperfine coupling constants values, which are finely tuned and significantly higher than those observed in biological systems, to detect effects at these high frequencies under geomagnetic conditions \cite{Lee2014, Ban2013, Nair2024}. Additionally, when we varied the OMF from 0 to 5 $\mu$T, the relevant range for the experiments under consideration, we found that the impact of ultra-high radio frequencies on radical pair systems remains negligible even at the estimated upper range amplitude, further questioning the radical pair mechanism's applicability in explaining experimentally observed effects on ROS production.

Therefore, while the RPM can well explain static magnetic field effects on ROS, it cannot account for the effects of telecommunication-frequency radiation on reactive oxygen species. Note that our modelling assumptions, such as utilizing a single nucleus H5 and disregarding dipole-dipole and exchange interactions, may have led to an overestimation of the effects observed \cite{efimova2008role, kattnig2017sensitivity, nohr2017determination, babcock2021radical, Luo2023}. This potential overestimation suggests that the actual impact of ultra-high radio frequencies on the radical pair system could be even lower, further emphasizing the significant limitation of the radical pair mechanism in explaining observed effects within biological systems. Given these limitations, it seems plausible that these effects may be more attributable to the electrical component of the electromagnetic field in the context of telecommunication devices. Our work highlights the need for a broader investigation to account for the observed effects.

\section*{Data availability}
Data and computational analysis are available from the corresponding author on reasonable request.

\bibliography{bibliography}

\section*{Acknowledgement}
The authors would like to thank Rishabh for his valuable comments and insights. This work was supported by the Natural Sciences and Engineering Research Council through its Discovery Grant program and the Alliance Quantum Consortia grant 'Quantum Enhanced Sensing and Imaging (QuEnSI)', as well as the National Research Council of Canada through its Quantum Sensing Challenge Program.

\section*{Author contribution}
H.Z.-H. and C.S. conceived the project; O.T. conducted the investigation and performed the modelling and calculations with assistance from H.Z.-H. and C.S.; O.T. wrote the paper with feedback from H.Z.-H and C.S.

\section*{Competing interests}
The authors declare no competing interests.

\end{document}